\documentclass[twocolumn,showpacs,amsmath,amssymb]{revtex4}
\usepackage{graphicx}

\begin{document}
\title{Time-bin modulated polarization-entangled biphotons from cavity-enhanced 
down-conversion}
\date{\today}
\author{Christopher E. Kuklewicz}
\email{chrisk@alum.mit.edu}
\author{Franco N. C. Wong}
\author{Jeffrey H. Shapiro}
\affiliation{Massachusetts Institute of Technology}
\begin{abstract}
  We have generated a new type of biphoton state by cavity-enhanced
  down-conversion in a type-II phase-matched, periodically-poled
  KTiOPO$_4$ (PPKTP) crystal.  By introducing a weak intracavity
  birefringence, the polarization-entangled output was modulated
  between the singlet and triplet states according to the arrival-time
  difference of the signal and idler photons.  This cavity-enhanced
  biphoton source is spectrally bright, yielding a single-mode
  fiber-coupled coincidence rate of 0.7\,pairs/s per mW of pump power
  per MHz of down-conversion bandwidth.  Its novel biphoton behavior
  may be utilized in sensitive measurements of weak intracavity
  birefringence.
\end{abstract}
\maketitle

Spontaneous parametric down-conversion (SPDC) is the principal source
for polarization-entangled photon pairs \cite{SPDC}, but its $\sim$THz
bandwidth makes it ill-suited for coupling to the $\sim$MHz bandwidth
of an atomic absorption line.  What is needed, to make such coupling
efficient, is a bright, narrowband source of polarization-entangled
photons \cite{ShapiroWong}.  Ou and Lu \cite{Ou99} used an optical
cavity to increase the efficiency and narrow the bandwidth of a
continuous-wave (cw) type-I phase-matched down-converter.  Wang \em et
al\/\rm.\@ \cite{Wang04} did the same for a pair of cw type-I
down-converters, one rotated by 90$^\circ$, in a ring cavity to
produce polarization-entangled photons.  We will report the first
cavity-enhanced operation of a cw type-II down-converter, resulting in
a spectrally bright, narrowband source of frequency-degenerate
polarization-entangled photon pairs.  More importantly, by controlling
a weak intracavity birefringence, our source generates a new type of
biphoton state whose polarization-entangled output is modulated
between the singlet and triplet states according to the arrival-time
difference between the signal and idler photons.  

Consider conventional cw SPDC in a collinear configuration that is
type-II phase matched at frequency degeneracy.  The post-selected
biphoton state emerging from a 50-50 beam splitter placed after the
usual timing-compensation crystal is then $|\psi\rangle =
(|H\rangle_1|V\rangle_2 + e^{i\phi}|V\rangle_1|H\rangle_2)/\sqrt{2}$,
in the horizontal-vertical ($H$-$V$) basis, where the subscripts label
the beam splitter output ports.  Ordinarily we have $\phi = 0$, so
$|\psi\rangle$ is a triplet, but a half-wave plate in one of the
output paths can transform $|\psi\rangle$ to a singlet by making $\phi
= \pi$.  The situation becomes more complicated---and more
interesting---when that down-converter is embedded in a single-ended
cavity that resonates the signal and idler.  If the signal and idler
photons resulting from down-conversion of a pump photon emerge from
the output coupler after the same number of roundtrips within the
cavity, then they yield a triplet for the post-selected biphoton
state.  However, the times at which these photons leave the cavity may
differ by integer multiples of the cavity roundtrip time, $\tau_c$.
Taking $\tau_c$ as a natural time-bin unit for the system, we have
that the biphoton state associated with a photon pair whose
arrival-time difference is $m\tau_c$ is $|\psi\rangle =
(|H\rangle_1|V\rangle_2 + e^{im\phi}|V\rangle_1|H\rangle_2)/\sqrt{2}$,
where $\phi$ is the roundtrip cavity birefringence.  By tuning the
cavity birefringence to achieve $\phi \neq 0$---something that is
easily done with type-II phase matching---we get a new type of
biphoton, which exhibits time-bin modulated polarization entanglement.
When $\phi = \pi$ and $m$ is even, this biphoton is a triplet; when
$\phi = \pi$ and $m$ is odd, it is a singlet.  Before describing our
experimental work, we shall provide a more precise characterization of
cavity-enhanced type-II SPDC.

\begin{figure}
  \centering\includegraphics[width=3.25in]{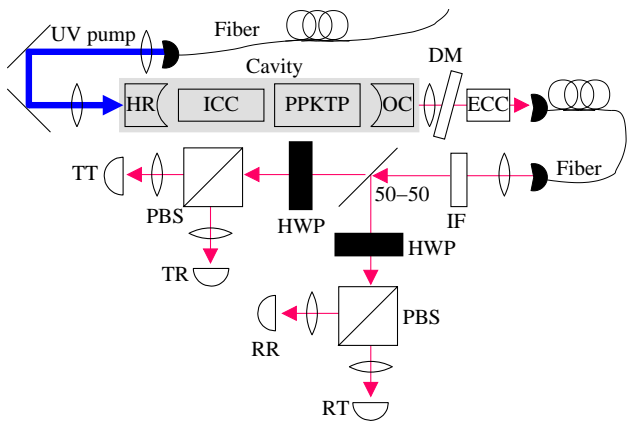}
  \caption{\label{fig:layout-cc-pp}(Color online) Cavity-enhanced SPDC
    setup. HR, high reflector; ICC, intracavity compensating crystal;
    ECC, external compensating crystal; OC, output coupler; HWP
    half-wave plate; PBS polarizing beam splitter; DM, dichroic
    mirror; IF, interference filter; TT, TR, RT, and RR, single-photon
    detectors.}
\end{figure}

Consider the cw down-conversion configuration shown in
Fig.~\ref{fig:layout-cc-pp}.  A length-$L$ KTP intracavity
compensating crystal (ICC) and a length-$L$ PPKTP crystal are
contained inside a single-ended optical cavity, and a length-$L/2$ KTP
external compensating crystal (ECC) is employed outside the cavity.
The cavity mirrors do not reflect the frequency-$\omega_P$ pump, but
they do resonate the frequency-degenerate signal and idler.  The pump
propagates along the $x$-axes of all three crystals. The $y$-axis of
the PPKTP crystal is horizontal, and the $y$-axes of the compensating
crystals are vertical.  Without the cavity mirrors, ICC has no effect;
with the cavity mirrors, ICC provides timing compensation for photons
that perform one or more roundtrips before exiting the
cavity.

Our experiments measured statistics of the time difference, $\tau$,
between photon detections made in the $H$-$V$ or the $\pm 45^\circ$
polarization bases.  The $H$-$V$ coincidence rate is $R_{HV}(\tau) =
C|K^{(p)}_{SI}(\tau)|^2$ in terms of the phase-sensitive correlation
of the signal and idler, $K_{SI}^{(p)}(\tau)$, and a proportionality
constant $C$, and the $\pm 45^\circ$ rate is $R_{\pm 45}(\tau)$ =
$C|K^{(p)}_{SI}(\tau) -K^{(p)}_{SI}(-\tau)|^2/4.$ Furthermore, we have
\cite{Kuklewicz05} $K^{(p)}_{SI}(\tau) \propto \sum_{m =
  -\infty}^\infty\sum_{\ell = 0}^\infty R^{\ell + |m|/2}e^{im\phi/2}
{\rm box}_{\tau_0}(\tau - m\tau_c)$, at a signal-idler double
resonance, where $R$ is the output coupler's reflectivity.  Also,
${\rm box}_{\tau_0}(\tau) = 1$ for $|\tau|\le \tau_0/2\equiv |\Delta
k'|L/2$ and zero otherwise, which is the time-domain manifestation of
the phase-matching function, with $\Delta k'$ being the frequency
derivative of the phase mismatch.  In our experiment, $\tau_0 \approx
3.5$\,ps, $\tau_c \approx 826$\,ps, and $\phi$ was controlled by
temperature tuning of the ICC\@.  Setting $R = 0$ reduces
$K_{SI}^{(p)}(\tau)$ to the SPDC result, i.e., the $\ell = m = 0$
term.  For $R >0$, the $m\neq 0$ terms are due to correlations between
signal and idler photons that came from the same pump photon but
exited the output coupler with the signal photon having taken $m$ more
cavity roundtrips than the idler did and incurring $m$ times the
cavity birefringence.  Because the different box functions in
$K_{SI}^{(p)}(\tau)$ are nonoverlapping in our experiment, we find
that $R_{HV}(\tau) = \sum_{m=-\infty}^\infty R_{HV}^{(m)}(\tau)$ and
$R_{\pm 45}(\tau) = \sum_{|m|=0}^\infty \sin^2(m\phi/2)
R_{HV}^{(|m|)}(\tau)$, where $R_{HV}^{(m)}(\tau)$ is the ${\rm
  box}_{\tau_0}(\tau-m\tau_c)$ term in the $H$-$V$
coincidence rate.

Our experimental setup is shown in Fig. \ref{fig:layout-cc-pp}.  The
intracavity compensating crystal was unpoled KTP polished to a length
within $\pm\,1\mu$m of the $\sim$9.7-mm-long PPKTP crystal, as
verified by birefringence measurements made with a tunable laser. Each
was placed on its own thermoelectric cooler for independent
temperature control with 0.01$^\circ$C stability.  The PPKTP crystal
was tuned to frequency degeneracy for SPDC, and the ICC was tuned to
control the cavity birefringence.  The PPKTP, ICC, and ECC crystals
were antireflection coated at 795\,nm and 397.5\,nm.  The ICC and
PPKTP were placed inside an optical cavity formed by an input mirror,
coated for high reflection at 795\,nm and an output coupler that was
coated for 92\% reflection at 795\,nm; both mirrors were also high
transmission at 397\,nm.  The mirrors had 50\,mm radii of curvature
and were separated by 104.9\,mm, which yielded a $\sim$1.21\,GHz free
spectral range ($\tau_c \approx 826$\,ps). The output coupler was
mounted on a piezoelectric transducer for cavity-length control.

We used a cw external-cavity diode laser (Toptica) as the pump source,
operated near 397.5\,nm.  The pump light was sent through 2\,m of
single-mode fiber (StockerYale) that served as a spatial filter, and
$\sim$1\,mW of power was mode matched into the cavity.  Most of the
pump light passed through the cavity and was then diverted from the
output path by the dichroic mirror (DM).  The cavity finesse at
795\,nm was measured to be $\sim$55, corresponding to 11\% loss per
roundtrip, which was consistent with the 92\% reflectivity of the
output coupler and the $\sim$0.3\% loss per surface for the crystals.
A $\sim$5\,mm KTP external compensating crystal was placed outside of
the cavity before the aspheric lens that coupled the down-converted
light into a single-mode fiber equipped with polarization-control
paddles.  The resonantly generated down-converted light had a well
defined spatial mode that allowed efficient mode matching into a
single-mode fiber.  After the fiber there was a 1\,nm interference
filter (IF) and a 50-50 beam splitter. The transmitted arm contained a
half-wave plate (HWP) set at $\theta_T$ and a polarizing beam splitter
(PBS) whose transmitted and reflected paths led to single-photon
counters TT and TR, respectively.  The reflected path from the 50-50
beam splitter contained another HWP (set at $\theta_R$) and a PBS with
two detectors (RT and RR) in its output paths.  The single-photon
detectors were all PerkinElmer silicon avalanche photodiodes.  Weak
pump-beam reflections ($\sim$17\%) at the cavity mirrors created a
weak pump modulation as the cavity length was swept.

We measured the coincidence rates between the TT and TR detectors and
between the RT and RT detectors.  TT and TR detections were also used
to provide start and stop signals for collecting
arrival-time-difference histograms for comparison with the predicted
behavior of $R_{HV}(\tau)$ and $R_{\pm 45}(\tau)$. The RT and RR
measurements were always made with $\theta_R=0$, corresponding to the
$H$-$V$ basis (aligned to the PPKTP's $y$ and $z$ axes).  Thus they
monitored the maximum coincidence rate to check that the
down-conversion remained consistent during data collection.  The TT-TR
coincidence rate and arrival-time difference histogram were measured
either with $\theta_T=0$ (for the $H$-$V$ basis) or with
$\theta_T=\pi/8$ (for the $\pm 45^\circ$ basis).  Data were collected
while the cavity length was swept through one free spectral range
(FSR) using a 40-s-period triangular waveform.  The
arrival-time-difference histograms at a signal-idler double resonance,
shown in the figures below, come from 16 minutes of data integration
using 38.3\,ps time bins.  Had the cavity length been locked at a
double resonance, the data collection time would have been
significantly reduced.

\begin{figure}
  \centering\includegraphics[width=3.25in]{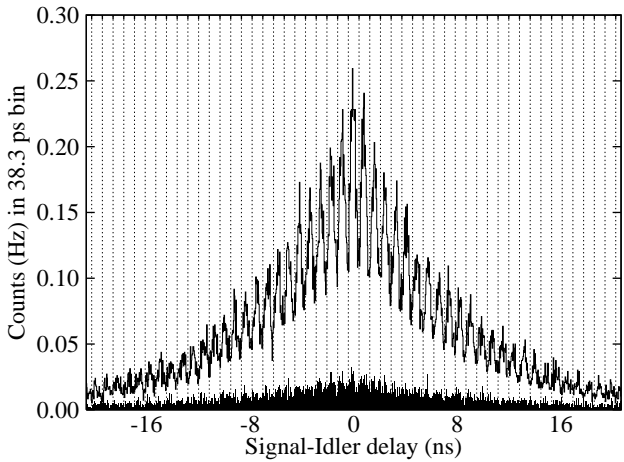}
  \caption{\label{fig:hist-good}Arrival-time-difference histograms
    averaged over 16 minutes.  Upper curve is for the $H$-$V$ basis.
    The lower filled curve is for the $\pm 45^\circ$ basis.}
\end{figure}

We measured the cavity transmission of a $\sim$795\,nm probe beam and
tuned the temperature of the ICC to simultaneously resonate both
polarizations as the cavity length was swept, after which the probe
laser was removed and SPDC coincidence counting was performed.  The
resulting TT-TR data are shown in Fig.~\ref{fig:hist-good},
corresponding to the case of zero birefringence, $\phi=0$.  The upper
curve is the arrival-time-difference histogram for the $H$-$V$ basis.
Its peaks are separated by the $\sim$826\,ps cavity roundtrip time
$\tau_c$, and broadened---from their 3.5\,ps theoretical width---by
the 350\,ps time jitter of our single-photon counters.  Its highest
($\tau = 0$) peak was due to signal-idler pairs that exited the cavity
after the same number of roundtrips.  The filled curve in
Fig.~\ref{fig:hist-good} is the arrival-time-difference histogram for
the $\pm 45^\circ$-basis.  Figure~\ref{fig:hist-good-ratio} fill~A
shows the ratio of $\pm 45^\circ$-basis coincidences, summed over the
peak for each $m$ value, to the corresponding sum for $H$-$V$-basis
coincidences, after a small background-count correction has been made
(0.014\,Hz/bin for $H$-$V$ and 0.009\,Hz/bin for $\pm45^\circ$).  The
ratio of the total of the central 41 peaks in the $\pm 45^\circ$
histogram to that for the $H$-$V$ histogram was 0.131.  This
corresponds to a quantum-interference fringe visibility of 76.8\%,
indicating that the SPDC output is in a polarization-entangled triplet
state \cite{Kuklewicz04}.  Much of the loss from ideal visibility can
be attributed to the reflected pump light, as discussed below.  The
slight curvature seen in Fig.~\ref{fig:hist-good-ratio} fill~A can be
attributed to temperature fluctuations of the ICC that were equivalent
to an offset of 0.004$^\circ$\,C from the ideal value.

By varying the ICC temperature, while keeping the PPKTP temperature
fixed, we were able to control the cavity birefringence precisely.  At
zero birefringence, the effective lengths of the ICC and PPKTP were
the same.  By varying the ICC temperature, the effective lengths could
be tuned to achieve $\phi = 2\pi T/T_{2\pi}$, where $T$ is the ICC
temperature relative to the zero-birefringence temperature, and
$T_{2\pi}$ is the temperature shift required to yield $\phi = 2\pi$.
Using a 795\,nm probe laser, we measured $T_{2\pi} = 4.5^\circ$C\@.
Thus the post-selected biphoton from our cavity-enhanced
down-converter should change from a triplet, which occurs at zero
time-bin difference ($m = 0$), to a singlet when $m T = T_{2\pi}/2$.

\begin{figure}
  \begin{center}
    \includegraphics[width=3.25in]{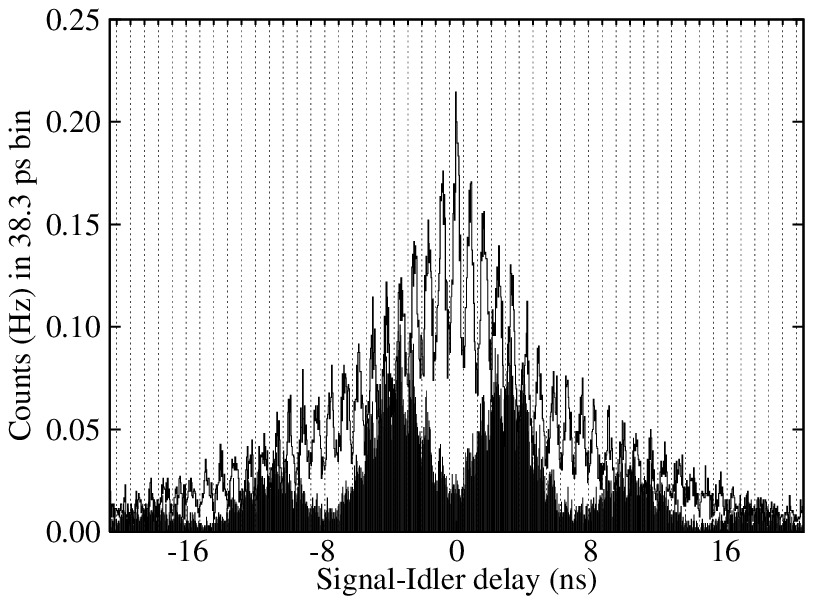}
  \end{center}
  \caption{\label{fig:h-hist}Similar to Fig.~\ref{fig:hist-good}
    with temperature detuned by 0.53$^\circ$C}
\end{figure}

Data were collected for a series of ICC temperatures, so that the
signal and idler resonated at different cavity lengths.
Figure~\ref{fig:h-hist} shows the results for $T \approx
0.53^\circ$C\@.  The upper curve in this figure is the $H$-$V$
histogram, and the filled curve is the $\pm 45^\circ$ histogram.  As
in Fig. \ref{fig:hist-good}, the central ($T =0$) peak in the $H$-$V$
histogram is suppressed in the $\pm 45^\circ$ histogram.  For $T\neq
0$, however, the $\pm 45^\circ$ coincidence rate shows the $\sin^2$
oscillation expected from our $R_{\pm 45}(\tau)$ expression, with a
period in $m$ of 8.90 roundtrips (fit to the data).  Thus the SPDC
output was modulated between triplet and singlet signatures, with
intermediate behavior occurring at roundtrip offsets in between these
extremes. Their histogram ratio, shown in
Fig.~\ref{fig:hist-good-ratio} curve~B, clearly indicates the periodic
change of the output state from triplet (ratio $\sim 0.2$) to singlet
(ratio $\sim 0.8$).

\begin{figure}
  \centering\includegraphics[width=3.25in]{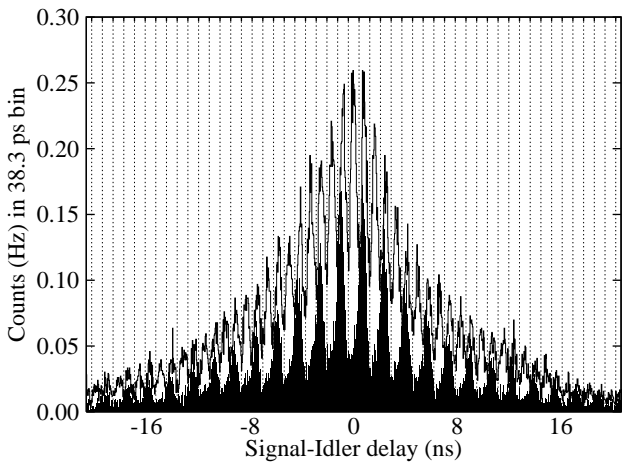}
  \caption{\label{fig:i-hist}Similar to Fig.~\ref{fig:hist-good}
    with the temperature detuned by 2.26$^\circ$C.}
\end{figure}

\begin{figure}
  \centering\includegraphics[width=3.25in]{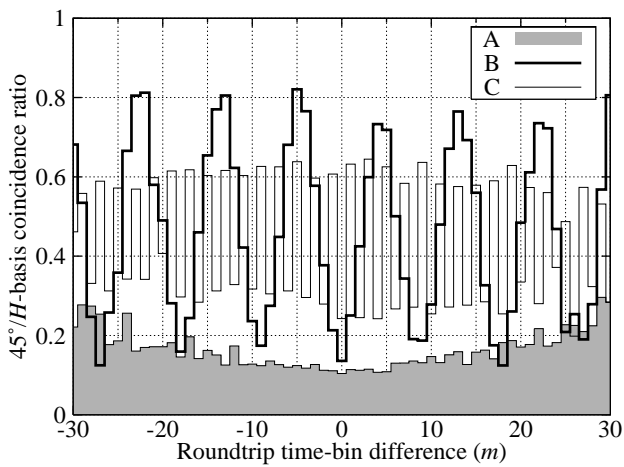}
  \caption{\label{fig:hist-good-ratio}Ratio of histograms from
    Figs.~\ref{fig:hist-good} (fill A), \ref{fig:h-hist} (curve B),
    and \ref{fig:i-hist} (curve C) versus roundtrip time-bin
    difference $m$.}
\end{figure}

Triplet behavior recurred whenever the temperature was detuned by a
multiple of 4.5$^\circ$C.  The data for Fig.~\ref{fig:i-hist} were
taken at $T = 2.26$$^\circ$C with $\phi = \pi$.  Now, all the
even-roundtrip peaks are suppressed and all the odd-roundtrip peaks
are maxima.  The ratio alternates between a maximum and minimum---see
Fig.~\ref{fig:hist-good-ratio} curve~C---in agreement with the
predicted period of 2 for this temperature detuning. The odd-roundtrip
peaks are exhibiting polarization singlet behavior.  The reduced
contrast was due to detector time jitter, which causes the counts from
one peak to spill into time bins associated with adjacent peaks.  This
effect also reduced the contrast in Fig.~\ref{fig:hist-good-ratio}
curve~B.

The $\sim$17\% pump reflection from the output coupler lead to
down-conversion in the backward pump path in Fig.~1.  However, the ECC
had the wrong orientation for pairs generated by this backward pump,
thus they always produce a $\pm 45^\circ$-basis to $H$-$V$-basis ratio
of 1/2, which reduced our measured quantum-interference visibility.
We rotated the ECC by 90$^\circ$, so that the forward-pump pairs have
a 1/2 ratio and the backward pairs exhibit quantum interference.  With
this arrangement the $\pm 45^\circ$-basis to $H$-$V$-basis ratio for
the $T=0$ peak increased from 12\%, for the normal-ECC orientation, to
43\%, for the rotated-ECC orientation, indicating that 90\% of the
down-conversion was producing interference either forward or backward.
Taking into account the ECC's 5~mm length not being exactly half the
length of the 9.7\,mm PPKTP crystal, this is expected to be 93.2\% for
a matched ECC\@.  The depths of the forward and backward dips imply
that 18.6\% of the pump was reflected, in agreement of the 17\%
estimate from the transmission measurements.  The remaining 6.8\%
contrast loss comes from unknown sources of signal-idler
distinguishability.

The peak TT-TR coincidence rate, during the cavity sweep, for the
light collected by the single-mode fiber was 2000\,pairs/s per mW of
pump power.  Because of the 50-50 beam splitter, this is 25\% of the
pairs from the fiber and 50\% of the coincidence rate between the
transmitted and reflected sides of the beam splitter.  This rate is
higher than the 300\,pairs/(s-mW) obtained from this PPKTP crystal
without cavity enhancement and without fiber coupling
\cite{Kuklewicz04}.  From the 280\,GHz phase-matching bandwidth,
1.21\,GHz FSR, and cavity finesse of 55, the FWHM of the central
frequency peaks should be $\sim$22\,MHz and should contain about 1/250
of the total output.  It follows that our experiment produced 0.7
polarization-entangled pairs/s per mW of pump power per MHz of
bandwidth at frequency degeneracy.  These are pairs from a single
fiber-coupled spatial mode, and are post-selected after a 50/50 beam
splitter.  In comparison, the single-pass source in Ref. \cite{Kuklewicz04}
produced only 0.001\,pairs/(s-mW-MHz) and the sources in Refs.
\cite{Kuklewicz04,Marco04} produced 0.003\,pairs/(s-mW-MHz) and
0.014\,pairs/(s-mW-MHz), respectively.  The type-I cavity-enhanced
source in Ref. \cite{Wang04} used a KbNO${}_3$ crystal and produced
0.12\,pairs/(s-mW-MHz).

In summary, we have demonstrated cavity-enhanced type-II phase-matched
SPDC producing spectrally bright, fiber-coupled polarization-entangled
biphotons that are suitable for efficient coupling to narrowband
atomic absorption lines.  We have generated a novel biphoton state of
time-bin modulated polarization entanglement that can be controlled by
adjusting the cavity birefringence.  The pure triplet biphoton state
can only be obtained at zero birefringence.  This unique property
suggests that it may be potentially useful for ultrasensitive
detection of weak intracavity birefringence.  For example, one may
place a normally isotropic material under thermal stress or
piezoelectric strain in a zero-birefringence cavity-enhanced biphoton
source.  Any stress-induced material birefringence would induce a
coincidence-rate modulation signature that yields the amount of the
material birefringence.

\begin{acknowledgments}
  This work was supported by a DoD MURI program under ARO-administered
  Grant No.\ DAAD-19-00-1-0177.
\end{acknowledgments}

\end{document}